\documentclass[12pt,oneside]{article}
\input epsf
\def\singlespace {\smallskipamount=3.75pt plus1pt
minus1pt
                  \medskipamount=7.5pt plus2pt
minus2pt
                  \bigskipamount=15pt plus4pt minus4pt
                  \normalbaselineskip=15pt plus0pt
minus0pt
                  \normallineskip=1pt
                  \normallineskiplimit=0pt
                  \jot=3.75pt 
                  {\def\smallskip
{\vskip\smallskipamount}}
                  {\def\medskip  
{\vskip\medskipamount}}
                  {\def\bigskip  
{\vskip\bigskipamount}}
                  {\setbox\strutbox=\hbox{\vrule 
                    height10.5pt depth4.5pt width
0pt}}
                  \parskip 7.5pt
                  \normalbaselines}

\def\doublespace {\smallskipamount=7.5pt plus2pt
minus2pt
                  \medskipamount=15pt plus4pt minus4pt
                  \bigskipamount=30pt plus8pt minus8pt
                  \normalbaselineskip=30pt plus0pt
minus0pt
                  \normallineskip=2pt
                  \normallineskiplimit=0pt
                  \jot=7.5pt
                  {\def\smallskip
{\vskip\smallskipamount}}
                  {\def\medskip  
{\vskip\medskipamount}}
                  {\def\bigskip  
{\vskip\bigskipamount}}
                  {\setbox\strutbox=\hbox{\vrule 
                    height21.0pt depth9.0pt width
0pt}}
                  \parskip 15.0pt
                  \normalbaselines}

\def\be{\begin{equation}}
\def\ee{\end{equation}}
\def\bea{\begin{eqnarray}}
\def\eea{\end{eqnarray}}

\def\sect #1{\setcounter{equation}{0}}

\begin{document}
\singlespace
%%\doublespace
%\begin{center}
\title{\Large{A Tolman-Bondi-Lemaitre Cell-Model for the
Universe and Gravitational
Collapse }}
%\end{center}
\vspace{1.0in}
\vspace{12pt}

\author{A. Chamorro${}^{1}$\thanks{wtpchbea@lg.ehu.es}, S. S. 
Deshingkar${}^{2}$\thanks{shrir@relativity.tifr.res.in}, 
I. H. Dwivedi ${}^{2}$${}^{,3}$\thanks{dwivedi@relativity.tifr.res.in} and
P. S. Joshi${}^{2}$\thanks{psj@tifr.res.in} \\
%\vspace{0.4in}
\\
${}^{1}$Departamento de Fisica Teorica, 
Universidad del Pais Vasco,\\ 
Apdo. 644, E-48080 Bilbao, Spain. \\
%%\vspace{0.4in}
\\
${}^{2}$Tata Institute of Fundamental Research, \\
Homi Bhabha Road, Mumbai 400005, INDIA \\
%\vspace{0.4in}
\\
${}^{3}$Permenent address: 17, Ballabh Vihar,\\
Dayalbagh, AGRA, INDIA \\}

\maketitle

\newpage

%\vspace{1.3in}

\begin{abstract} 
A piecewise Tolman-Bondi-Lemaitre (TBL) cell-model for
the universe incorporating local collapsing and expanding inhomogeneities
is presented here. The cell-model is made up of TBL underdense and
overdense spherical regions surrounded by an intermediate region of TBL
shells embedded in an expanding universe. The cell-model generalizes the
Friedmann as well as Einstein-Straus swiss-cheese models and presents a
number of advantages over other models, and in particular the time
evolution of the cosmological inhomogeneities is now incorporated within
the scheme. Important problem of gravitational collapse of a massive dust
cloud, such as a cluster of galaxies or even a massive star, in such a
cosmological background is examined. It is shown that the collapsing local
inhomogeneities in an expanding universe could result in either a black
hole, or a naked singularity, depending on the nature of the set of
initial data which consists of the matter distribution and the velocities
of the collapsing shells in the cloud at the initial epoch from which the
collapse commences.
 
\end{abstract}

%\newpage
%\doublespace

\section{Introduction}

The conventional cosmological scenarios are based on the Friedmann-
Robertson-Walker (FRW) solutions of the Einstein equations. These models
have the advantage of being simple, because the universe has been assumed
to be isotropic and homogeneous on large enough scales of the order
of 300 Mpc and higher. It is thus possible to use these models for making
a number of predictions on the large scale structure and evolution of the
universe. The disadvantage, however, is that one is no longer able to take
into account in an exact manner the various observed inhomogeneities
present in the universe, such as the development and evolution of
structures such as voids, or local collapsing inhomogeneities such as a
gravitationally collapsing massive cloud in a cosmological background,
which may represent a sufficiently isolated  collapsing massive star which has 
exhausted its nuclear fuel, or a star cluster, or even a 
cluster of galaxies evolving dynamically.
 
The important unsolved problem is thus that of taking
into account and modeling the various inhomogeneities, and their
evolution in the real universe. We propose here an idealized cell-model for the
universe which attempts this task. The universe is no longer
assumed to be made up of a uniform and homogeneous matter continuum, but
is taken to be consisting of various elementary building blocks embedded in 
a uniform background. In fact, a complete 
solution of Einstein equations for a matter continuum consisting of 
inhomogeneous dust is available, as given by the so called 
Tolman-Bondi-Lemaitre (TBL) models \cite{TBL}. The FRW models form a special
class of this more general class of models. Using earlier work of
Chamorro \cite{C1}, and Bonnor and Chamorro \cite{BC1}, it is then shown how a 
model  for the evolution of the inhomogeneities in the universe can now be
constructed. One of the advantages over former swiss-cheese models is that our 
picture allows a much richer and diverse structure of the universe as 
different structures may exist within other structures, all of them 
ultimately embedded in an otherwise expanding universe. The universe 
not only has inhomogeneities, but these inhomogeneities may not be uniform 
in structure. The model presented here, besides generalizing previous 
swiss-cheese models, makes an attempt to incorporate the inhomogeneities
suitably.

We then turn our attention to the problem of gravitational collapse
of a massive cloud in such a cosmological background.  The gravitational 
collapse scenarios involving the collapse of a compact body have 
been explored in quite some detail in recent years, particularly 
in the context of the cosmic censorship hypothesis (see e.g. Joshi \cite{PSJG}, 
and references there in). The scenario that has been developing 
from these studies is that for various forms of matter such as dust, 
perfect fluids, radiation collapse etc. both black holes and naked 
singularities develop as the end state of collapse depending on the nature of
the initial data from which the collapse commences. In fact, it turns
out that  sets of initial data of matter configurations, in
terms of densities and velocity profiles prescribed at the onset of 
collapse, may yield evolutions within the context of general relativity 
which would lead to either a black hole or 
a naked singularity, independently of the form of matter or the 
equation of state used.

Most of these models have so far been  analyzed within an asymptotically 
flat background, which is best suited for phenomena involving isolated 
bodies.  However, in  widely accepted cosmological models describing
the present scenario of the expanding 
universe, spacetime curvatures are non-vanishing throughout 
the universe, and therefore such a picture of collapse could not be 
taken to represent a real gravitational collapse of cosmic relevance. 
On the other hand, considerations of a collapsing inhomogeneity in 
an expanding universe would represent a more realistic physical 
situation. It is therefore of interest to study the nature of 
gravitational collapse in such expanding environments. That way one 
could find out if the introduction of a cosmological background 
could possibly affect the nature and outcome of the collapsing local 
inhomogeneity. Our consideration here show however that the incorporation 
of a cosmological background does not change the earlier conclusions
on gravitational collapse  qualitatively. That is, 
depending on the nature of initial data which consists of the 
density distribution and velocity profiles  of the collapsing shells
in the cloud,  both black holes and naked singularities would arise.

The plan of the paper is as follows. In Section 2 the cell-model for the 
universe is constructed, describing its basic building blocks which will 
account for the elementary inhomogeneities. In Section 3, 
we consider the gravitational collapse problem in such a cosmological
background. Section 4 examines the issue of global visibility
of such singularities for faraway observers in the universe. 
In the concluding Section 5 we discuss the overall scenario and conclusions, 
and the possibilities of generalizing these results further.

\section { A Tolman-Bondi-Lemaitre cell-model for the
universe }

In line with current ideas about the cell structure of the universe 
we construct here a model for the universe made up of TBL, and
in particular, Friedmann underdense and overdense spherical regions 
surrounded by compensating thick TBL shells. As we shall see, ultimately
all inhomogeneities will be embedded in an expanding Friedmann background of 
either positive, zero or negative spatial curvature.  In construction
of such a cell-model we follow and extend ideas contained 
in Chamorro \cite{C1}, where models of voids in elliptic Friedmann 
universe were presented.

Throughout the paper we use Einstein field equations $G_{ab}=
-kT_{ab}$ with vanishing cosmological constant, and the cosmic fluid is 
taken as the dust matter with the stress-energy tensor given by 
$T_{ab}=\rho u_au_b$. All elementary inhomogeneities would be 
assumed to be spherically symmetric. The field equations for dust matter have
already been solved and the solution is the TBL
metric given by

\be
ds^2= -dt^2 +{{R'^2}\over{1+f}}dr^2 + R^2d\Omega^2,
\ee

\be
\rho={F'\over R^2R'},\quad \dot R^2=f+{F\over R}
\ee

Here $f(r)$ and $F(r)$ are arbitrary functions of $r$
only, and are interpreted
as the energy and mass functions. The energy
density is $\rho$, while (.) and ($'$) 
denote partial differentiation with respect to $t$ and
$r$. 

\begin{figure}[h]
\parbox[b]{7.88cm}
{
\epsfxsize=7.85cm
\epsfbox{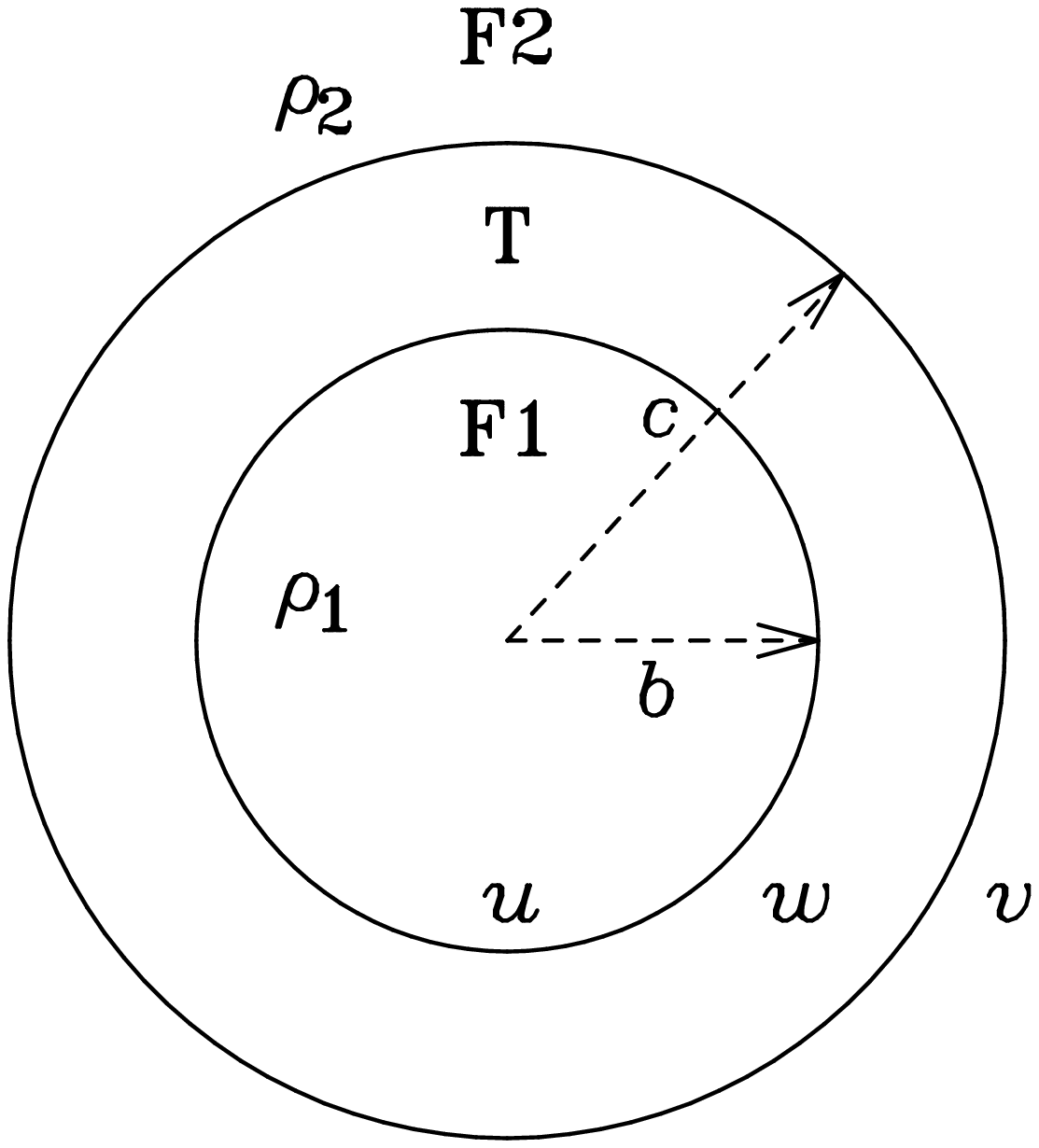}
}
\caption{
}
\end{figure}

Before discussing the basic building blocks of the
cell-model in detail, which consist of various types of local inhomogeneities, 
let us first consider how a single inhomogeneity is inserted in our
cell-model in general.  The local inhomogeneity (see Fig. 1) 
is described by a TBL underdense or overdense region F1 (as the case may be )
and is surrounded by a compensating region T of a thick
TBL shell which in turn is embedded in the background cosmology of region F2. 
The metric and the density functions in the three regions are
given as
\be
ds^2= -dt^2 +{{R_1'^2}\over{1+f_1}}dr^2 +
R_1^2d\Omega^2,\quad r<b,
\quad \rho_1={F_1'\over R_1^2R_1'}
\ee
\be
ds^2= -dt^2 +{{R'^2}\over{1+f}}dr^2 +
R^2d\Omega^2,\quad b<r<c,\quad
\rho={F'\over R^2R'}
\ee
\be
ds^2= -dt^2 +{{R_2'^2}\over{1+f_2}}dr^2 +
R_2^2d\Omega^2,\quad c<r,\quad
\rho_2={F_2'\over R_2^2R_2'}
\ee

The solution of Einstein equations in region F1 is
given as,
\be 
R_{1}={F_{1}\over - 2 f_{1}}(1-\cos u),\quad t-a_1(r)=
k_1(u-\sin u), \quad r<b 
\ee
where $k_1$ is given by,
\be 
k_1={F_{1}\over 2(-f_{1})^{3/2}}
\ee
For the region F1 we have $f(r)<0$, $r\ge 0, 0\le
\theta \le \pi,
0\le \phi \le 2\pi$ and  $0<u<2\pi$, $t>a(r)$.

Note that the
shell labeled by $r$ is initially singular at $t=a(r)$
for $u=0$(big bang),
and since region F1
is elliptic (i.e. $f<0$) in case it is collapsing later
on it also becomes 
singular in the future at the time,
\be
t=t_s(r) =a(r) + {\pi F\over (-f)^{3/2}}.
\ee

For the regions T and F2, depending upon whether the
region is elliptic,
parabolic or hyperbolic (i.e. $f(r)<0, =0, >0$), the
solutions
have the following form,
\be
R_{i}= -{F_{i}\over 2f_{i}}(1-\cos {\it v}), \quad t
-a_{i}(r) =
{F_{i}\over 2(-f_{i})^{3/2}}
({\it v}-\sin {\it v}) \; for \; f<0
\ee
\be
R_{i}={\bigg (}{9F_{i}\over 4}{\bigg )}^{1/3}
[t-a_i(r)]^{2/3} \; for \;
f=0
\ee
\be
R_{i}= -{F_{i}\over 2f_{i}}(1-\cosh {\it v}), \quad t
-a_i(r) = -{F_{i}\over 2f_{i}^{3/2}}
({\it v}-\sinh {\it v}) \; for \; f>0.
\ee
In the elliptic region ${\it v}$ takes values from $0$
to $2\pi$, while in the hyperbolic region it takes values from $0$ to
$\infty$. For the intermediate region T we would use the notation
$(R_i\equiv R, F_i\equiv F, f_i\equiv f, a_i\equiv a, {\it v}\equiv w)$ 
while for the cosmological background region F2 it shall be $(R\equiv R_2, 
F\equiv F_2, f\equiv f_2, a_i\equiv a, {\it v}\equiv v)$

If the spacetime as described in Fig 1 is to be considered a solution of the
field equations, the solution must satisfy the boundary conditions at $r=b$ and
$r=c$. The Darmois matching conditions at the boundaries $r=b$ and
$r=c$ (see Bonnor \& Vickers 1981\cite{BV}) require
$$F_{1}(b)=F(b),\quad f_{1}(b)=f(b)$$
$$F_{2}(c)=F(c),\quad f_{2}(c)=f(c)$$
\be R(b,t)=R_1(b,t), \quad R_2(c,t)=R(c,t) \ee

Thus the above conditions determine the
boundary values of the mass 
and energy functions $(F,f)$ of the intermediate
region T for a given
inner region F1 and outer cosmological region F2.
Note however that there may be discontinuities in the densities across
the shells $r = b, c$, as continuity of the functions $F', f'$ and $R'$ is not
required by (12).

The ratio of the average density of the
region F1, which is $\rho_1$, to the exterior Friedmann universe
$\rho_2$,
is given for elliptic universe by
\be 
{\rho_1\over \rho_2}={\gamma[v(c))]\over
\gamma[u(b)]}\left({t-a_{1}(b)
\over t-a_{2}(c)} \right)^2,
\ee
\be 
\gamma (v)={(1-\cos v)^3\over (v-\sin v)^2},
\ee
and for the hyperbolic case it is given by,
\be
{\rho_1\over \rho_2}= -{F(c)^2\over
F(b)^2}{f(b)^3\over f(c)^3}
{[\cosh v(c) -1]^3\over [1- \cos u(b)]^3},
\ee
where as in the parabolic case it is,
\be
{\rho_1\over \rho_2}= -{18f(b)^3\over F(b)^2}
{[t-a_{2}(c)]^2 \over
[1-\cos u(b)]^3}.
\ee

Note that if the outer Friedmann region is parabolic or hyperbolic then
in the region T we need to have a transition from $f<0$ to $f\ge 0$. This
can be achieved smoothly in the TBL model provided $f(r)$ 
and $F(r)$ be at least $C^2$ functions.
This is because though we write the solution in three different regions in
three different ways it is actually a one continuous function as
all the functions and derivatives from either side
of $f=0$ match at the transition point, i.e. $f=0$ \cite{NEW}. 
An equally important aspect of the solution, if the inhomogeneity is to be 
inserted smoothly through the intermediate region to the outside cosmological 
region F2, is that there be no shell-crossings. To avoid shell-crossings we 
must require for the mass function
$F(r)$, $f(r)$ and the singularity time $a(r)$ to satisfy Hellaby and
Lake's conditions as given in Table I of \cite{HL}.

We discuss now in detail the building blocks or
elementary 
inhomogeneities of the model. For the sake of clarity
we would first consider
these inhomogeneities in various situations in an
outside elliptic
exterior (i.e. $f_2(r)<0$).

2.1  Expanding void in expanding exterior : 

The solutions for the metric (1)
corresponding to the regions shown in figure 1 are, 

In region F1,
\be 
R_1= k_1 r(1- cos u),  \quad t+\epsilon = k_1(u - sin u)
\ee
$r<b$, $u< \pi$.

In region T,
\be
R= F(-2f)^{-1}(1- cos w), \quad
(w - sin w) = 2(-f)^{3/2}F^{-1}(t-a) \ee
$b < r <c$, $w < \pi$ .

In region F2,
$$
R_2= k_2 r(1- cos v), \quad t = k_1(v - sin v)
$$
and $r>c$, $v<\pi$.  

%\begin{figure}[h]
%\parbox[b]{7.99cm}
%{
%\epsfxsize=7.95cm
%\epsfbox{t1a.ps}
%}
\  \  \
%\parbox[b]{7.99cm}
%{
%\epsfxsize=7.95cm
%\epsfbox{t1b.ps}
%}
%\caption{
%I is the underdense region. The outer Friedmann
%expanding region is II. 
%In between lies the Tolman compensating zone T.}
%\end{figure}

The inequalities for the angular parameters $u,v$ and
$w$ at the present time 
$t=t_0$ are needed to ensure expansion in the three regions. The
Darmois matching 
conditions are enforced at $r=b$ and $r=c$ .  
In this case a non-simultaneous big-bang (NSBB) is
required if the ratio 
$\rho_1/\rho_2$ is to be less than 0.6 (see for
details Chamorro \cite{C1}).

2.2  Expanding void in contracting exterior :  

As in the previous case, but
now $v>\pi$ at $t_0$. Then $v >\pi>u$, and since $d\gamma
/dv<0$ one has ${\gamma[v]\over
\gamma[u]}<1$;  
that according to (13)  makes it possible to have
$\rho_1/\rho_2<1$ taking suitable values for $a_1(b)$ and $a_2(c)$. 
Then by 
using the Hellaby and Lake conditions for no shell-crossings, 
it can be seen
(as in reference [2]) that the avoidance of these
singularities requires
having $R'<0$  in some region of T. The general
appearance of the profiles
of the functions $-f$, $F/2k_1b$ and $a$ versus $r$ in
the intermediate
TBL zone ($b,c$) may be the same as that of Fig. 2
in the same reference.
However, there is no need of a NSBB to get
$\rho_1/\rho_2$ as small as wanted, 
provided we choose the ratio $k_1/k_2$ large enough.
Therefore, one can  also 
set $a(r)=0$ to have a simultaneous big
bang (SBB) for this
building block.

2.3  Contracting overdense region in expanding exterior:

In this instance one needs $u(t_o)>\pi>v(t_o)$, that
leads to
$\rho_1/\rho_2 >1$ for all t and SBB by just taking
$k_1$ conveniently 
smaller than $k_2$. 
The intermediate zone T may be described by
choosing a function 
$p(r)\ge 0$
for $r \epsilon (b,c) $ large enough and setting $F'/F
= 3f'/2f +p(r)$
to have $F'\ge 0$, that together with $a' =0$ (SBB in
T) leads to
$R'>0$ and no shell-crossings (see ref. [2]).

2.4 Summary of the former and the other possible building blocks :

There are a total of eight elementary inhomogeneities 
corresponding to the following characteristics :

\begin{tabular}{|l|p{0.8in}|p{1.10in}|p{3.50in}|}
\hline
1. & $\rho_1/\rho_2 <1$    & F1:expansion, F2: expansion &
SBB:   $k_1>k_2$, NSBB: $k_1 \stackrel {{ >}}{<} k_2$ \\
\hline

2. &  $\rho_1/\rho_2< 1$ &  F1:expansion, F2:contraction
& SBB: 
$k_1 >k_2$, NSBB: not required \\
\hline

3.&  $\rho_1/\rho_2>1$ & F1:contraction, F2:expansion &  SBB
: $k_1<k_2$.\\
\hline

4.&   $\rho_1/\rho_2>1$  & F1:expansion, F2:expansion &SBB :
$k_1<k_2$, $a_1=0$, $a_2>0$\\
\hline

5.&  $\rho_1/\rho_2 <1$ & F1:contraction, F2:expansion & 
Required NSBB:
        $a >0$; profiles for f, F and $a$ as in case 1 for no shell-crossings
nor surface layers.\\
\hline

6.&  $\rho_1/\rho_2 <1$ & F1:contraction, F2:contraction & 
Essentially reducible to cases
        3 or 5 as the outermost F-background must be
expanding.\\
\hline

7.&    $\rho_1/\rho_2>1$ & F1:expansion, F2:contraction & 
Requires NSBB, but essentially reduces to cases 3 or 5 for same
reason as in case 6\\
\hline

8.&    $\rho_1/\rho_2>1$ & F1:contraction, F2:contraction & 
Essentially reduces
to cases 3 or 5 again, depending on whether
$\rho_2/\rho_0 \ge 1$.
$\rho_0$ is the density of the outermost F-background. \\
\hline

\end{tabular}

2.5   Final considerations: 

The Darmois matching conditions together with
Einstein equations yield 
that $1 / 2 F(c)$ can be interpreted as the active
gravitational mass 
within the shell of radius $r=c$. That happens to
coincide with the mass 
function introduced by Cahill and McVittie \cite{CMV}. Thus
one might construct a cell-like dust model for the universe 
by inserting elementary inhomogeneities of the 
kind considered above in an expanding Friedmann
background as depicted in  
Figure 2. The following interesting features of the
model should be noticed :

\begin{figure}[h]
\parbox[b]{7.99cm}
{
\epsfxsize=7.95cm
\epsfbox{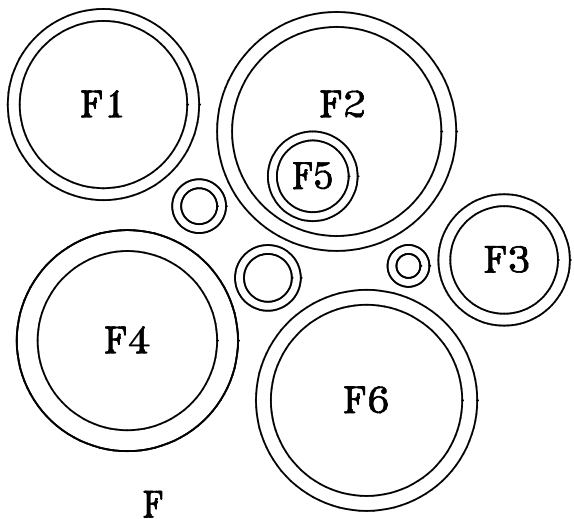}
}
\  \  \
%\parbox[b]{7.99cm}
%{
%\epsfxsize=7.95cm
%\epsfbox{t1b.ps}
%}
\caption{
F is the outermost expanding Friedmann background.  F1, F2,.... stand 
for the voids and overdense regions, expanding or contracting. They are 
surrounded by their corresponding compensating TBL zones.
}
\end{figure}

a) It allows to put inhomogeneities such as F5 within
other inhomogeneities (when the inhomogeneities contain homogeneous and
isotropic parts) in 
contradistinction to the traditional swiss-cheese
models $`$a la Einstein-Strauss.

b) The geometrical center of each inhomogeneity may be
taken as its center of
mass. All such centers may be viewed as fundamental
particles of the cosmic 
fluid, moving exactly as if they were particles of the
exterior F-background
within which all the inhomogeneities are embedded.

c) Thus the overall average expansion of this model
would be described by the
dynamics of the exterior Friedmann background.

d) This provides a cell  model for the expanding
universe where the
average problem of Einstein equation: $G_{\mu
\nu}(<g>) \ne -<T_{\mu \nu}>$,
finds a simple solution.

Though we have discussed here the case of
positive curvature
Friedmann universe, the results can be easily
generalized to the case
when the outside Friedmann universe is of zero or
negative curvature. In 
these cases, if $f_1(r) < 0$, the intermediate TBL region has to take us
from positive 
curvature to zero or negative curvature. As it was mentioned above
this
is possible because though we
write the TBL solution for the three cases in three
different forms, it is actually
a one continuous solution if the change over is
smooth. That is, it is enough that the
energy and mass functions in T be $C^2$ for the solutions of the
Einstein equations to
exist. The case $f(r) \geq 0$ can be dealt with without the incommodity
of shell-crossings by taking negative curvature solutions for 
part or all of Region T and for F2.

\section{Gravitational Collapse and Singularity
Formation}

The occurrence of a physical phenomena, as 
predicted by the theory, is possible only if the
conditions assumed under 
which the proposed models are derived are physically
realistic, and are
realized in nature. In this context, since the
universe is widely believed 
to be described by an expanding cosmological model at
its present stage, 
it is quite natural to ask whether all the conclusions
drawn from 
consideration of collapse scenarios of isolated bodies
in an asymptotically flat background would
hold if the cosmological background is taken into
consideration. 
That is, whether a collapsing massive cloud in an expanding
universe would give rise 
to the same conclusions regarding black holes or naked
singularities
formation. We therefore investigate this issue here
with the
aim of examining the possible formation of a naked singularity,
or a black hole,
during the collapse of 
a dense body in an otherwise cosmological background,
to see whether 
the background imposes any constraints on the occurence
of these phenomena.

In fact, it is known that a massive star in the
universe does not have 
sharp boundaries. The star or the compact body has a
core which is superdense 
and an outer layer and crust which is less dense,
which is further surrounded 
by a cloud of much lower density. In a realistic
collapse situation, 
there is always a strong possibility that the core of
the star, 
being superdense and massive, may undergo a continued
collapse,
while the outer layers of the star may be blown up far
away. Thus, in 
real situations, the core of the star would be
collapsing while
there would be an intermediate region consisting of a
much less dense cloud
which could be expanding in a smeared-out expanding universe.
Certainly, such considerations apply even more
effectively in the
case of possible collapse of very large aggregations
of matter in the 
universe, such as a supercluster of galaxies or the
great attractor.
The cell-model of the universe described in the
earlier section 
thus becomes quite relevant to a real physical
scenario.

In the context of gravitational collapse, spherical
dust clouds in general 
relativity have been studied quite extensively \cite{NTBL}. The
prescription of matter 
as pressureless dust could perhaps be regarded as
somewhat idealized. 
However, some authors have considered dust as a good
approximation of the 
form of the matter in the final stages of collapse
(see e.g. Penrose \cite{Penrose1}). 
In any case, the study 
of dust models has led not only to
important new advances
and insights into various aspects of the gravitational
collapse theory, but 
have also laid the foundation of black hole
physics. We therefore 
consider here the problem of a superdense region of
dust in an expanding 
Friedmann model, and see if the formation of a naked
singularity or a black 
hole would occur as it did in asymptotically flat exteriors.

We shall use here the cell-model formerly described and
contemplate
the situation of an overdense collapsing region
surrounded by an 
underdense intermediate region, in the exterior
background of 
an expanding Friedmann universe. 

The basic idea, in order to
examine the existence
or otherwise of naked singularities forming as end
state of gravitational 
collapse, is to investigate the structure of families
of non-spacelike geodesics of the spacetime to find if
there are such 
families which terminate in the past at the
singularity, and in
future they are outgoing, reaching a faraway observer.
If there are
no such families the collapse ends in a black hole,
otherwise we have a naked singularity in the
spacetime. While outgoing
families of non-spacelike geodesics can be treated in
general, it is
enough for the present purpose to examine outgoing
null geodesics
radiating away from the singularity. Using the
notation, 
\be
X = {R \over r^{\alpha}}, \quad \Lambda =
{F\over r^{\alpha}} 
\ee
the equation for radial null geodesics can be written
as,
\be
{dR\over d(r^{\alpha})} = \left[ 1- \sqrt{{f + {\Lambda\over
X}}\over 1+f} \right]
{R'\over \alpha r^{\alpha-1}} \equiv U(X,r^{\alpha}). 
\ee
We use $r^{\alpha}$ instead of r for convenience in order to examine
the structure of the singularity.  The exact value
of the constant $\alpha\ge 1$ depends on the different
initial density
and velocity distributions for the collapsing cloud.
We first
examine the geodesic equation in Region I in order to
understand the
nature of the singularity which occurs during the
collapse of this
overdense region. The outgoing null trajectories, if
any, then can be
traced into Region T and finally to the expanding
Region F2.
The point $r^{\alpha}=0, R=0$ is a singularity of the
above 
differential equation in Region I. 
In order to understand the nature of outgoing
families, it is necessary to examine the nature of
this singularity.

If future directed null geodesics do terminate in the
past at the
singularity $R=0, r^{\alpha}=0$ with a definite tangent, this
is determined by
the limiting value of $X= R/r^{\alpha}$ at $R=0,r^{\alpha}=0$. In such a
case,
\be
X_0 = \lim_{R\to0,r^{\alpha}\to0} {R\over r^{\alpha}}= 
\lim_{R\to0,r^{\alpha}\to0}
{dR\over d(r^{\alpha})} 
=\lim_{R\to0,r^{\alpha}\to0} U= U(X_0,0).
\ee
If a real positive value of $X_0$ satisfies the above
equation,
then the singularity could be naked.
The above equation has been analyzed in
detail. It turns out that the necessary and sufficient
condition for the central shell-focusing singularity
to be naked,
at least locally, is that equation (21) admits a real,
positive root
$X=X_0$. In fact, for the case of TBL models, it is
possible to
write this equation explicitly as an algebraic
equation in terms
of the basic initial data parameters $F$ and $f$ \cite{INI},
and it follows that the central singularity, occurring
at the time $t=t_s(0)$
where $t_s(r)$ denotes the future singularity curve
$R(t,r)=0$, 
is naked under quite
general conditions which essentially depend on the
behavior of $F$ and
$f$ near the center. In particular, for the case when
\be
\rho'(0) \ne0, \Theta(0) >0
\ee
where $\Theta(r)= r {t_s}'(r)\sqrt {h(r)}/r^{3(\alpha
-1)/2}$, 
the central singularity is
locally naked. When $\rho'(0)=0$, a similar condition
can be given in
terms of $\rho^{''}(0)$. In all these cases, the
global nakedness
depends on the global behavior of the functions $F$
and $f$ away from
$r=0$. We discuss this issue in the next section.

We note that for the local nakedness of the
singularity, the behavior
of the functions (subject to $C^2$ differentiability)
is allowed
to be completely arbitrary for $r>0$. Thus for
collapse in a 
cosmological background, that is, for a collapsing region I,
 that need not be necessarily homogeneous, the
behavior of $F_1$ and $f_1$ at its boundary is
not restricted by the local nakedness condition.

So, for a given set of density and velocity
profiles, if the
collapse produces a naked singularity, then a suitable
intermediate region
T with the inner boundary at $r=r_1$ can exist
such that outgoing null
geodesics would cross that boundary and
might finally escape
to a distant external observer.

\section{Global Visibility of  Singularities}

Amongst the various versions of cosmic censorship
available in the literature,
the strong cosmic censorship hypothesis does not allow
the singularities
to be even locally naked. The spacetime must then be
globally hyperbolic.
Hence the case of the existence of a real positive root to
equation (21) is a
counter-example to the strong cosmic censorship conjecture \cite{INI}.
However, one may take
the view that the singularities which are only locally
naked may not be of
much observational significance as they would not be
visible to observers
faraway in the universe, and as such the spacetime outside the
collapsing object may be
asymptotically predictable.  On the other hand, globally
naked singularities could
have observational significance, as they are visible
to an outside observer
faraway in spacetime. Thus, one may formulate a
weaker version of
censorship, whereby one allows the singularities to be
locally naked but
rules out global nakedness. 
We examine now this
issue of local versus global visibility of the
singularity for the models with cosmological background 
we have been considering. That will
provide insight concerning the possibility of
global visibility when we vary the initial data, 
as given by the initial velocity distributions of the cloud.
\begin{figure}[p]
\parbox[b]{6.88cm}
{
\epsfxsize=6.85cm
\epsfbox{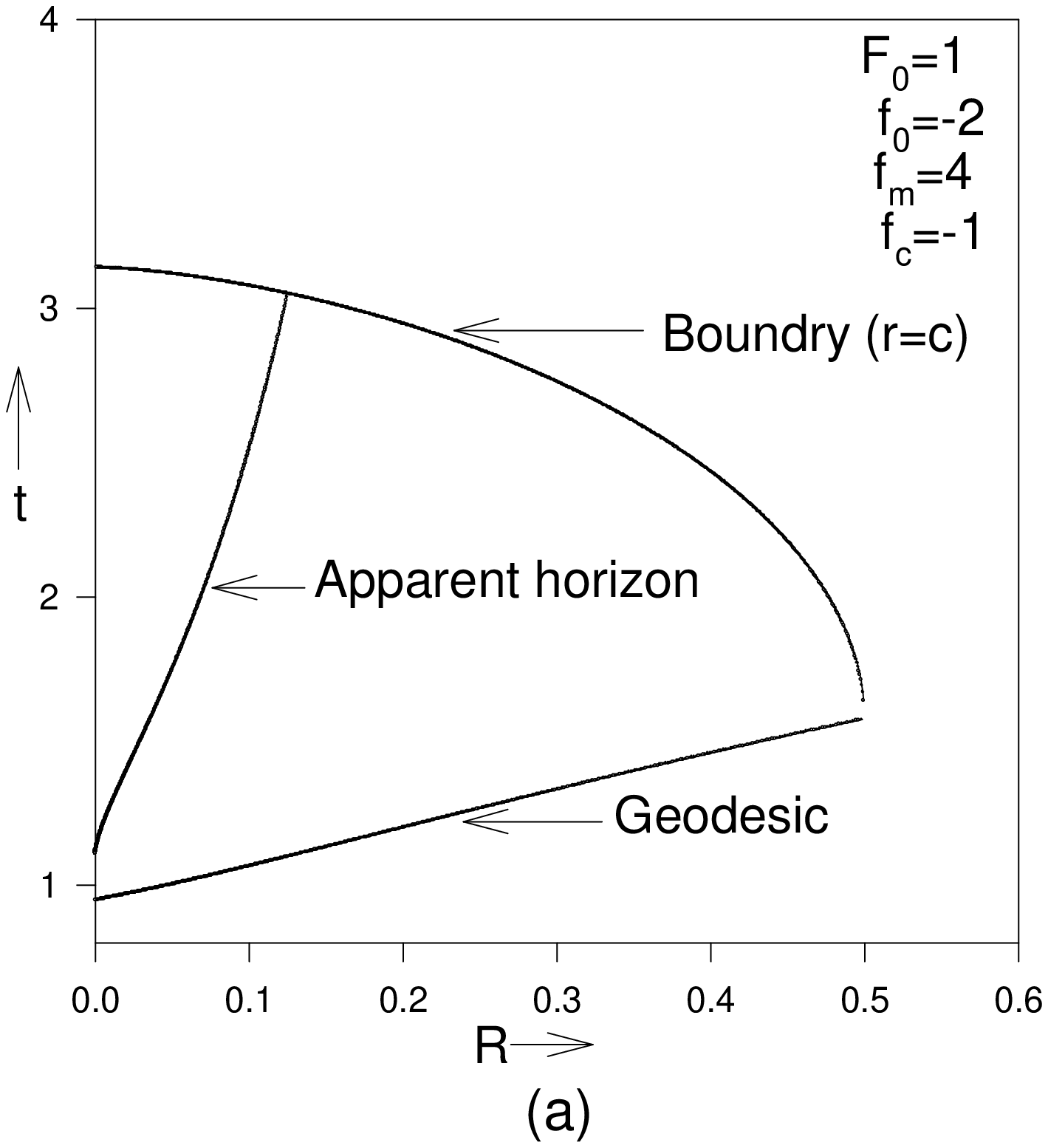}
}
\ \ \
\parbox[b]{6.88cm}
{
\epsfxsize=6.85cm
\epsfbox{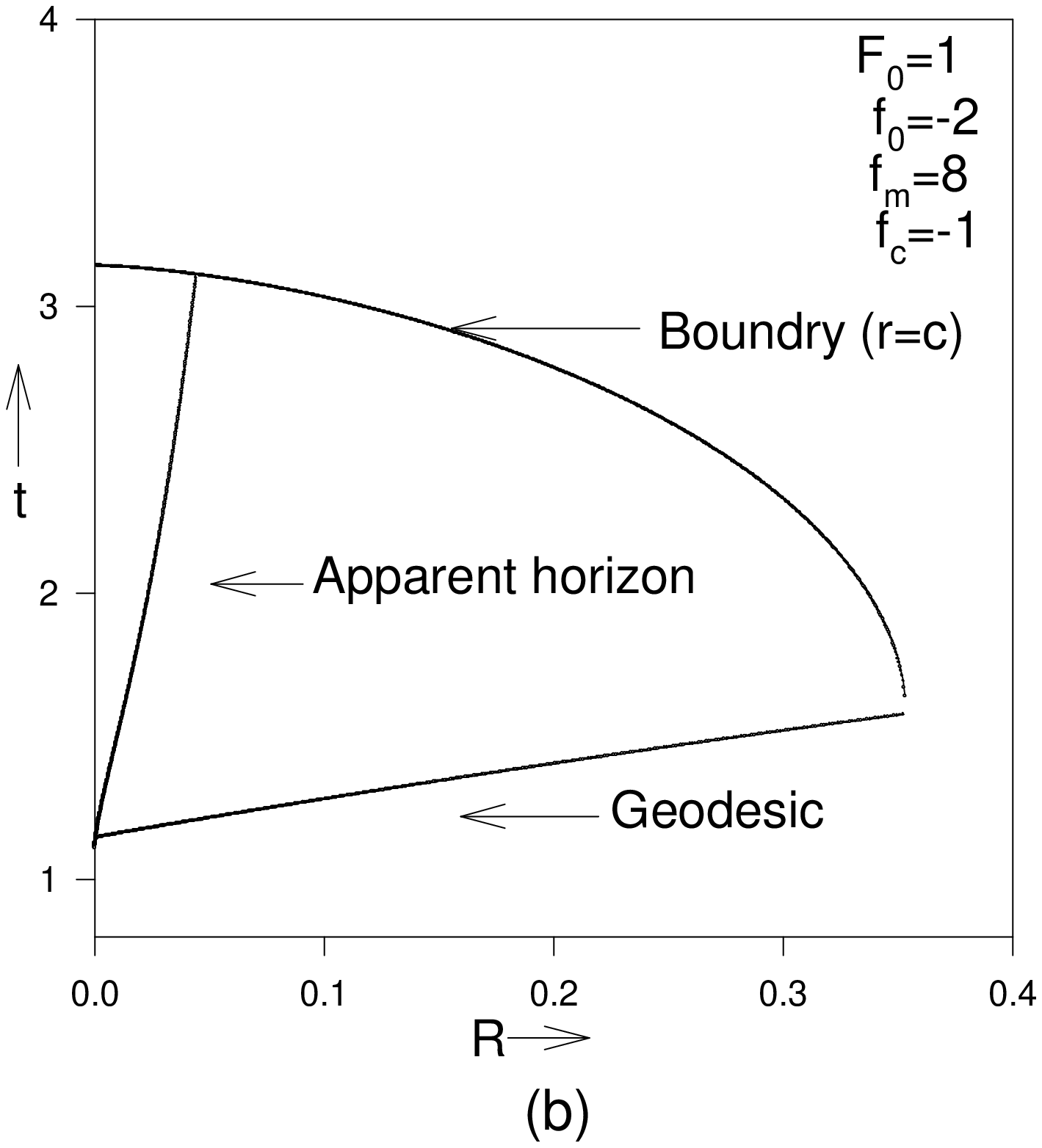}
}
\end{figure}
\vskip 1.5cm
\begin{figure}[p]
\ \ \
\parbox[b]{6.88cm}
{
\epsfxsize=6.85cm
\epsfbox{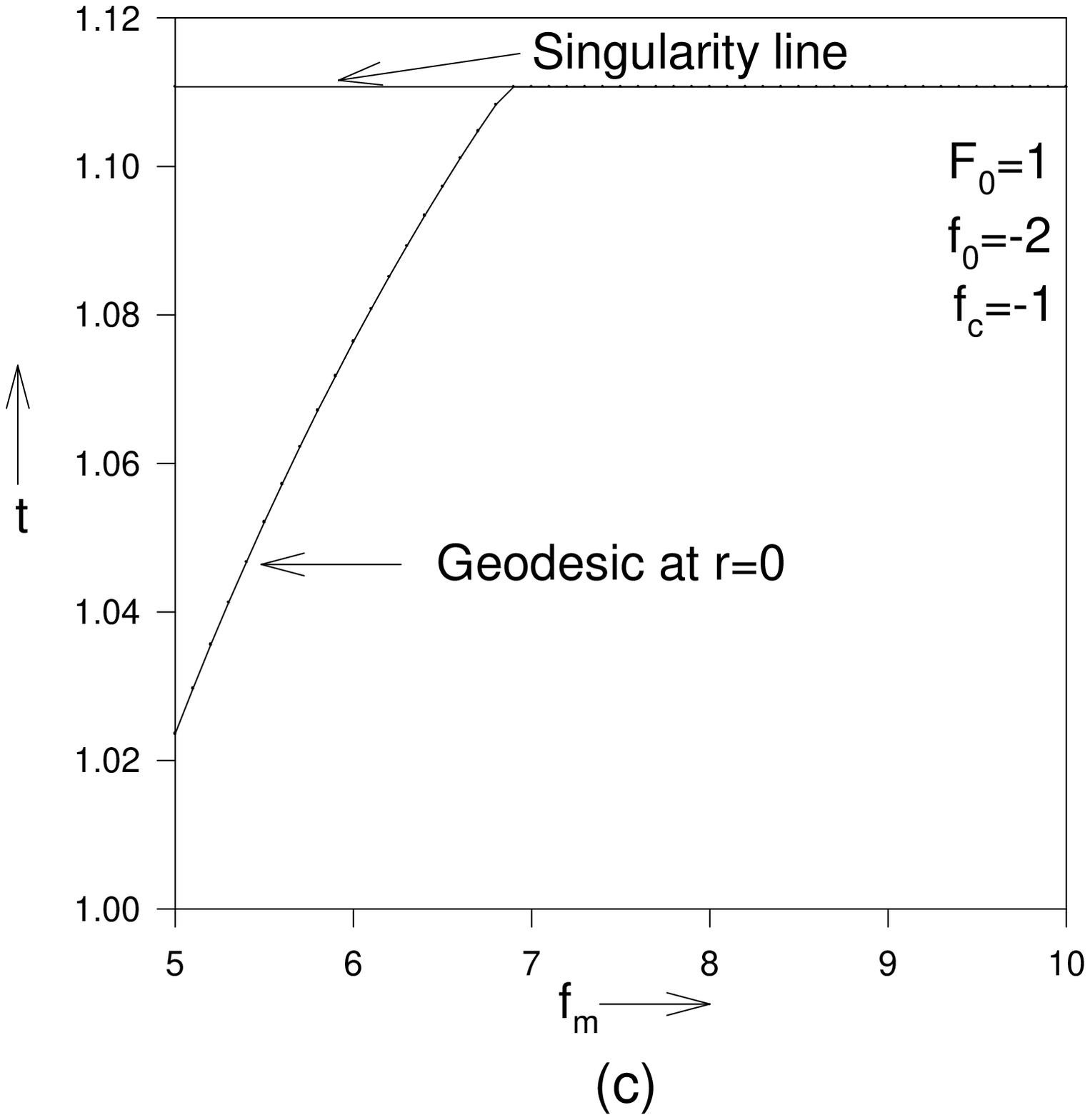}
}
\caption{                                                                       
The structure of singularity in the closed ($f<0$) universe case. For
the inner TBL collapsing core we take $f(r)= f_0r^2 + f_mr^4$ and $F=F_0r^3$.
In the cosmological region $f=-r^2$ and $F=F_0r^3$.
Here $f_0=-2$ and $F_0=1$, so the central singularity forms at
$t=t_0(0)=1.1$. (a) $f_m=4$ case. The event horizon meets the center $r=0$ below 
the singularity and no null trajectories escape to a distant observer.
The singularity is only locally naked.(b) $f_m=8$. The singularity is globally
naked.  There is a family of null geodesics from the singularity which meet
the boundary of cloud with $R>-F(c)/f(c)$.
(c) Graph showing transition from local nakedness to
global nakedness, by varying $f_m$.}
\end{figure}
                               
It is shown below that
both locally and globally visible singularities,
are possible within the scenario we have considered here,
depending on the initial
conditions. 
By varying suitable parameters in the regular initial data,
we examine the effect on the global
visibility of the singularity.
Here we study the case of an elliptic universe.
We take $a=0, F_1=F_2=r^3, f_2=-r^2$, and $f_1
= -2r^2 +f_m r^4 $ with $f_m$ such that $f_1<0$ within Region I, and
we can, in this case, match the inner TBL model (Region I, which consists 
of a collapsing core and an outer expanding region at the 
initial epoch) directly with the
outside Friedmann model (Region F2)
by satisfying the Darmois matching conditions at the
boundary $r=c$, which implies,

\be
c= \sqrt{1/f_m}
\ee

We also need that $c<1$, i.e. $f_m>1$. We shall say that the
singularity is globally
visible if the null geodesics coming out of the
singularity meet the $r=c$ shell before it starts recollapsing. 
As we vary the parameter $f_m$, we see a transition from a 
locally naked 
to a globally naked singularity at $f_m=6.75$. 
Examples of naked singularities which are only locally naked, 
as well as those of globally naked singularity in a cosmological background,
and of the transition, are illustrated in Figure 3.

%        We also study the case of parabolic and
%hyperbolic universe
%in the similar way. For parabolic case the boundary is
%fixed at 
%$c= \sqrt{2/f_m} $ and for the hyperbolic case it is
%at $c= \sqrt{3/f_m}$.
%Note that in this case also for regularity we need
%$f_m>1$. In this case
%we call the singularity to be globally naked if the
%geodesic meet the
%matching surface, as in that case they will keep on
%going out forever. We see
%a transition  from local to global nakedness even in
%this cases also. The
%results are shown in figure 4 and 5.

\section{Concluding Remarks}

The issue of the end state of gravitational collapse in
physically 
reasonable configurations is most crucial to the
cosmic censorship 
hypothesis. In this context, the present work shows
that gravitational 
collapse of local inhomogeneities in an otherwise
expanding universe 
can give rise to naked (locally or globally) 
or covered singularities,
depending on 
the initial data of the matter cloud. These
conclusions are in line 
with the earlier conclusions on collapse of a star
against an 
asymptotically flat background. It thus appears that
singularities, 
both naked or covered, are in a way natural to general
relativity.

Our work here shows that naked singularities, or black
holes,
do occur as the end stage of collapse in a cosmological
background,
where the curvatures are non-vanishing throughout the
universe.
It follows that the assumption of asymptotic flatness
is not crucial 
to the formation of a covered or naked singularity (see also \cite{WMV} for
an example in Vaidya background).
Considerable amount of work has been done in black
hole physics, 
however black hole physics is heavily dependent on
the existence 
of asymptotic flatness, which in fact is not the situation
in the large scale physical
universe. Thus, questions arise such as wether or not some of the
results of black hole physics would be different
if considered 
in a cosmological background. The answer to these
questions is 
certainly important  as it would bring
black hole 
physics closer to  physical reality, and thus
brighten the 
chances of obtaining its observable features in our
universe.

We acknowledge the support from
University of Basque Country Grants UPV122.310-EB150/98, UPV172.310-G02/99
and the Spanish Ministry Grant PB96-0250.

\end{document}